\documentclass[a4paper,11pt]{article}
\usepackage{aaskaiid}
\usepackage[capitalise,noabbrev]{cleveref}
\usepackage{xspace}
\usepackage{amsmath}
\usepackage{aaskaiid,subfigure,orcidlink}

\newcommand{\camb}{\texttt{CAMB}\xspace}
\newcommand{\ccl}{\texttt{CCL}\xspace}

\newcommand{\halofit}{\texttt{halofit}\xspace}
\newcommand{\ofour}{O$_{\rm IV}$\xspace}

\newcommand{\lcdm}{\ensuremath{\Lambda}CDM\xspace}
\newcommand{\Omegam}{\ensuremath{\Omega_{\rm m}}\xspace}

\newcommand{\ska}{SKA-Mid AA4\xspace}

\def\review#1{{#1}}

\title{Weak Lensing with SKAO: Cosmic Shear Cosmology}
\ShortTitle{Weak Lensing}

\author[1]{Ian Harrison\orcidlink{0000-0002-4437-0770}}
\ShortName{Harrison et al.} 
\author[2-7]{Roberto Ingrao\orcidlink{0009-0009-6082-9818}}
\author[2,3,8,9]{Stefano Camera\orcidlink{0000-0003-3399-3574}}
\author[10]{\\Catherine Cress\orcidlink{0000-0003-3483-3951}}
\author[11]{Mamta Pandey-Pommier\orcidlink{0000-0001-5829-1099}}
\author[12,13,14]{Ziad Sakr\orcidlink{0000-0002-4823-3757}}
\author[15,16]{Cora Uhlemann\orcidlink{0000-0001-7831-1579}}

\affiliation[1]{School of Physics and Astronomy, Cardiff University, CF24 3AA, UK}
\emailAdd{harrisoni@cardiff.ac.uk}
\affiliation[2]{Dipartimento di Fisica, Universit\`a degli Studi di Torino, Via P.\ Giuria 1, 10125 Torino, Italy}
\affiliation[3]{INFN -- Istituto Nazionale di Fisica Nucleare, Sezione di Torino, Via P. Giuria 1, 10125 Torino, Italy}
\affiliation[4]{Dipartimento di Fisica, Sezione di Astronomia, Universit\`a di Trieste, Via Tiepolo 11, 34131 Trieste, Italy}
\affiliation[5]{INFN -- Istituto Nazionale di Fisica Nucleare, Sezione di Trieste, Via Valerio 2, 34127 Trieste, Italy}
\affiliation[6]{INAF -- Istituto Nazionale di Astrofisica, Osservatorio Astronomico di Trieste, Via Tiepolo 11, 34143 Trieste, Italy}
\affiliation[7]{IFPU -- Institute for Fundamental Physics of the Universe, Via Beirut 2, 34151 Trieste, Italy}
\affiliation[8]{INAF -- Istituto Nazionale di Astrofisica, Osservatorio Astrofisico di Torino, Strada Osservatorio 20, 10025 Pino Torinese, Italy}
\affiliation[9]{Department of Physics \& Astronomy, University of the Western Cape, 7535 Cape Town, South Africa}
\affiliation[10]{Department of Mathematical Sciences, University of South Africa, Florida Park, Roodepoort 1709, South Africa}
\affiliation[11]{Pole Scientific, University Catholic of Lyon, Campus Saint-Paul, 10 place des Archives 69288, Lyon Cedex 02, France}
\affiliation[12]{Instituto de Física Teórica UAM-CSIC, Campus de Cantoblanco, 28049 Madrid, Spain}
\affiliation[13]{Institut de Recherche en Astrophysique et Plan\'etologie (IRAP), Universit\'e de Toulouse, CNRS, UPS, CNES, 14 Av. Edouard Belin, 31400 Toulouse, France}
\affiliation[14]{Universit\'e St Joseph; Faculty of Sciences, Beirut, BP-11514, Lebanon}
\affiliation[15]{Fakultät für Physik, Universität Bielefeld, Postfach 100131, 33501 Bielefeld, Germany}
\affiliation[16]{School of Mathematics, Statistics and Physics, Newcastle University, Herschel Building, Newcastle-upon-Tyne, NE1 7RU, UK}

\abstract{We discuss the power of weak gravitational lensing surveys with the SKAO in constraining cosmological parameters and the properties of radio star-forming galaxy samples. As well as reviewing progress to date on cosmic shear in radio experiments, we show forecasts for parameter constraints using the Mid telescope both alone and in cross-correlation with contemporaneous optical surveys. By selecting a sample of resolved, high-redshift star-forming galaxies in Band 2, surveys with the AA4 configuration will be capable of measuring the growth of structure on large scales in the Universe through the effect of weak gravitational lensing on their shapes. Assuming the high fidelity reconstruction of such galaxy shapes to be possible, we find that SKAO will measure the $S_8$ structure formation parameter to a level of $5\%$ alone and $3\%$ in full combination with either LSST or the \emph{Euclid} satellite. These measurements will be highly important due to their radically different sensitivities to key weak lensing systematics, both instrumental and astrophysical, and as such provide a vital robustness test to a pillar of modern cosmological measurements. Radio surveys also provide unique and potentially game-changing information in the form of polarisation and galaxy kinematics, which allow the cleaner separation of lensing from intrinsic galaxy shapes and can increase statistical power by factors $\sim5$-$10$.}

\begin{document}
\maketitle

\section{Introduction}
Cosmic shear refers to the weak but statistically detectable change in the shapes of images of distant sources due to gravitational lensing. By measuring shapes of large numbers of distant galaxies, we can trace the large scale gravitational potential as a function of cosmic time, and hence the baryonic and dark sector physics which builds large scale structures in our Universe. From initial detections at the turn of the century \citep{Bacon:2000sy,Kaiser:2000if,vanWaerbeke:2000rm,Wittman:2000tc} cosmic shear using angular power spectra of optical and near-IR galaxy shapes has matured into a highly sensitive probe of cosmological parameters , providing low-redshift measurements which complement the high-redshift Cosmic Microwave Background (CMB) in testing cosmological models. Detecting the cosmic shear signal typically requires the measurement of shapes of $\gtrsim 1$ galaxy per square arcminute within a $\gtrsim 10\,$deg$^2$ survey. Such surveys will be possible with SKAO's Mid-frequency telescope \citep[as defined in][]{braun2019anticipatedperformancesquarekilometre}, and previous studies \citep{Brown:2015ucq,Bonaldi:2016lbd,Harrison:2016stv,SKA:2018ckk} have made the case for their ability to do cosmology competitive with the ``Stage-III'' optical surveys currently approaching completion (DES: \citealt{DES:2026mkc}; HSC: \citealt{Dalal:2023olq, Li:2023tui}; KiDS: \citealt{Wright:2025xka}). As well as being \emph{competitive} such surveys are also highly \emph{complementary}. Because of its statistical nature, the history of weak lensing cosmic shear cosmology has been characterised by a continuous interplay between increasing precision from better experiments and the need for greater modelling accuracy. Multiple experimental and astrophysical systematic effects can mimic the cosmic shear signal and lead to biased results and erroneous conclusions on the physics of  structure growth. In radio surveys instrumental systematics are radically different, with stochastic PSF uncertainty in ground-based optical experiments replaced by highly-deterministic interferometer dirty beams at high radio frequencies, albeit at the cost of added PSF complexity in the form of strong side-lobes. This means radio cosmic shear surveys can be used to either remove or calibrate otherwise troublesome systematics, whilst retaining overall constraining power, through cross-correlation with surveys in other wavebands \citep{Harrison:2016stv,Camera:2016owj,Kalaja:2024tsk}. \review{As well as instrumental systematics, radio surveys can mitigate astrophysical cosmic shear systematics by providing information from galaxy polarisation and kinematics measurements. These trace the unlensed shape of sources, which are otherwise unavailable from the total intensity image, as discussed in \cref{sec:systematics}.}  Radio star-forming galaxy samples are expected to extend to higher redshifts than typical optical samples, increasing the strength of the lensing signal and extending the redshift to which structure growth can be measured.

The cosmic shear observable considered here is typically combined with galaxy clustering and galaxy-galaxy lensing in so-called $3\times2$pt analyses which typically provide the best low redshift \review{$z \lesssim 2$} measurements of structure growth through the $S_8 = \sigma_8 (\Omegam / 0.3)^{0.5}$ parameter. We defer to \cite{Harrison01.2026.SKA} also in this volume for discussion of $3\times2$pt and further $N\times2$pt combinations involving SKAO cosmic shear.
 
In this chapter we focus on the cosmological precision which will be available from auto-spectra of SKA-Mid AA4 surveys and cross-correlation spectra with Stage-IV optical/nIR (\ofour) surveys (\cref{sec:aa4shear}). This represents an update (necessitated by changing experiment timelines) to previous forecasts which combined AA4-like SKA-Mid configurations with current Stage-III optical surveys, and `full SKA / SKA2'-like configurations with Stage-IV optical surveys.
\section{Importance of AA4}
\begin{figure}[h]
    \centering
	\includegraphics[width=0.6\columnwidth]{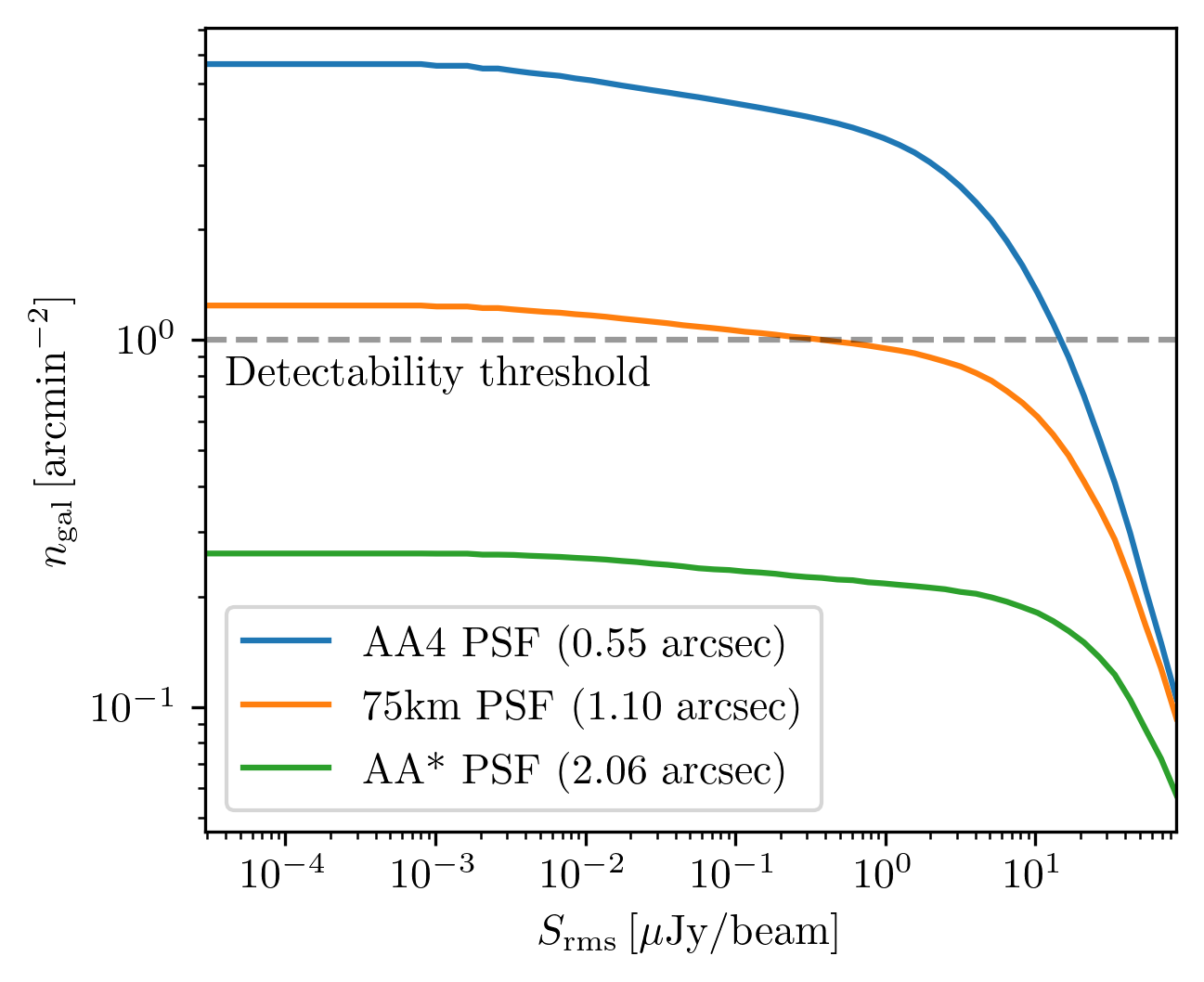}
    \caption{Sky number density of resolved galaxies in an SKA-Mid continuum survey, as a function of integration depth. In order to detect the cosmic shear signal a density of $1 \, \mathrm{gal}/\mathrm{arcmin}^{-2}$ is required. Though this is easily achieved with AA4 it is unachievable with AA* no matter the noise level reached. Moving the fixed number of AA* dishes to create longer 75$\,$km baselines recovers the ability to achieve the threshold number density.}
    \label{fig:aa4_aastar}
\end{figure}

Detecting the weak lensing cosmic shear signal requires the measurement of the shape (specifically the ellipticity) of source objects at redshifts $z\gtrsim0.5$. Star-forming galaxies are the preferred source as it is easiest to measure their ellipticity with the necessary accuracy. Observed and modelled source populations of star-forming galaxies emitted in the radio have an angular size distribution typically peaking around $1\,$arcsec, meaning a PSF size of this or smaller is required. Detailed simulations to optimise radio weak lensing surveys \citep{Bonaldi:2016lbd} show that the highest signal-to-noise on the shear signal, and the greatest cosmological information, is obtained from a survey operating in SKA-Mid Band 2 (centred at $1050\,$MHz), with a PSF size of $0.55\,$arcsec and integrating to a depth of $2.9\,\mu\mathrm{Jy/beam}$. This optimisation assumed an AA4-like array configuration, with maximum baselines up to $150\,$km. To investigate the relative performance of the AA$^{*}$ configuration we adopt the same procedure as \cite{Bonaldi:2016lbd} in cutting the T-RECS simulated source catalogue \cite{Bonaldi:2018xfm} such that sources are detected with SNR$>5$ and resolved with major axis size \review{larger} than the beam. For a detectable cosmic shear signal we require a source number density after these cuts of $>1$ galaxy arcmin$^{-2}$. As shown in \cref{fig:aa4_aastar}, no matter the integration depth this threshold of sky number density of resolved galaxies is never reached with the PSF size available from the AA$^{*}$ configuration in Band 2. For information we also show a case in which $75\,$km baselines are included in the SKA-Mid layout. A PSF resulting from this configuration would allow detection of weak lensing, albeit at a higher cost in terms of integration depth. This could be balanced by concentrating on smaller fields, with \cite{Bonaldi:2016lbd} showing that cosmological measurements can still be made down to total areas of $\sim 100\,\mathrm{deg}^2$, and cluster weak lensing \citep{Pandey-Pommier02.2026.SKA} remaining scientifically compelling on order of magnitude smaller fields. While strong lensing measurements probe the highly non-linear inner regions of gravitational potential wells, and cosmic shear is sensitive to the large-scale, linear regime of structure formation, weak lensing measurements of galaxy clusters provide a crucial intermediate measurement. They enable the reconstruction of mass density profiles over a broad radial range, from $\sim10\,\mathrm{kpc}$ in the cluster core to several $\mathrm{Mpc}$ in the outskirts, effectively bridging the gap between the small and large scale lensing regimes. Thus, joint analyses of strong and weak lensing within the same fields offer a powerful means of validating shape measurement algorithms and diagnosing systematic biases in shear estimation.
\section{Weak Lensing with Radio Shapes}
Whilst SKA-Mid surveys with AA4 will have the source properties necessary for cosmology with cosmic shear, there remain important methodological developments still to make. Because the shear signal is statistical, in that it requires averaging over many galaxy shapes, even small biases in shape measurement from the data can overwhelm the signal. In the optical there has been a long and ultimately successful programme of development to achieve the necessary precision and accuracy \citep[e.g.][]{Sheldon:2017szh}. Replicating this in radio interferometer data is an ongoing avenue of investigation, with promising methods which work both in the image and $uv$ plane. This topic is the subject of its own chapter in this volume \citep{Tripathi01.2026.SKA}. Here we discuss the current observational status of radio weak lensing given current methods, and in the subsequent forecasts in \cref{sec:aa4shear} we assume that this programme will be successful. \review{We also point out that the cross-correlation analyses advocated as a method of calibration in \cref{sec:systematics:calibration} can mitigate residual shape measurement systematics equally and jointly in  radio and optical surveys. We model such systematics in \cref{sec:systematics:forecasts:results} as randomly realised additive and multiplicative biases of the cosmic shear power spectrum \citep[a long-established treatment of cosmic shear systematics, e.g.][]{Amara:2007as}.}

\subsection{Observational Status}
The earliest attempt at a detection of radio weak lensing was in the VLA FIRST survey, in which \cite{Chang:2004ys} were able to apply a $uv$-plane shapelet method to measure galaxy ellipticities, resulting in a $3.0$-$3.6\sigma$ detection of an aperture lensing mass. Subsequent works using FIRST in combination with the optical SDSS survey \citep{Demetroullas:2015gsa,2018MNRAS.473..937D} also measured the cross-correlation power spectrum between the radio and optical surveys to a detection significance of $2.7\sigma$ and the stacked lensing of FIRST objects around SDSS galaxies at $10\sigma$.

Subsequent efforts have focused on JVLA observations of the COSMOS field, which contains a high number density of high redshift star forming galaxies, albeit in a relatively small $\sim 2\,$deg$^2$ field and with too low resolution to obtain shapes for many. \cite{Tunbridge:2016oio} analysed the $1.4\,$GHz VLA COSMOS images in conjunction with HST-ACS optical data to constrain the scatter between radio and optical source position angles, and \cite{Hillier:2018gdr} used $3\,$GHz JVLA data in the same field to detect a significant correlation between position angles across wavelengths, and set an upper limit on the cross-power spectrum of radio-optical cosmic shear.

To address the lack of available resolution from the JVLA, a number of studies have made use of the UK \emph{e}-MERLIN interferometer. Following \cite{Patel:2009rr}, who investigated shape correlations in a small number of MERLIN sources, the upgrade of the facility to \emph{e}-MERLIN enabled far greater depth on larger fields. The SuperCLASS \emph{e}-MERLIN legacy survey was the first set of observations specfically designed with a radio weak lensing detection in mind. The analysis of the first half of the SuperCLASS data \citep[DR1]{Harrison:2020zsv} resulted in an upper limit on the shear radio-optical cross-power spectrum, whilst demonstrating that the techniques \review{developed} should allow the (as of writing ongoing) analysis of the full data set to make a detection of the significant cluster lensing signal present in the field.

More recently the advent of reliable high fidelity imaging with the international LOFAR (iLOFAR) baselines has raised the prospect of measuring the ellipticities of star-forming galaxies at lower frequencies. \cite{Liu:2025sbv} use observations of the ELAIS-N1 field in both $\sim 150\,$MHz iLOFAR and optical HSC data to measure the shape covariance between the two bands, detecting a signficant correlation. They also forecast from the available source number densities in existing data that a survey of $\sim7\,$deg$^2$ in 3200 hours with iLOFAR would detect a cosmic shear signal at $>6\sigma$, which would represent a highly important stepping stone between existing observations and the $\mathcal{O}(1000)\,$deg$^2$ surveys assumed with SKA-Mid.

The prospect of a cosmologically informative cosmic shear survey in the radio has long been discussed \citep{Kamionkowski:1997mp,Blake:2004pb,Brown:2015ucq,Harrison:2016stv,Bonaldi:2016lbd,Camera:2016owj,Bull:2018cuw,Connor:2021tpa}, but relies on the presence of longer baselines than available in SKAO precursors, and the necessary access to low-level data products in order to allow for shape measurement \citep{Patel:2015cra,Harrison:2015lba}. Both these issues will be solved with the arrival of SKA-Mid AA4, allowing $\sim 5\%$ measurements of the $S_8$ parameter and the advent of multi-wavelength constraints robust to systematics.

\subsection{Utility as a Tool to Calibrate Cosmological Systematics}
\label{sec:systematics:calibration}
Throughout the history of cosmic shear cosmology there has been a continued interplay between statistical and systematic limitations on the cosmological information the probe can provide. As statistical samples of galaxies increase in size, the increasing available cosmological precision necessitates higher levels of control of systematics which can mimic the signal, be they instrumental (e.g. PSF mis-estimation), methodological (e.g. shape mis-modelling) or astrophysical (e.g. the intrinsic alignment of galaxy shapes). For a comprehensive account of these systematics, see e.g. \cite{Mandelbaum:2017jpr}. The extension of the understanding of these effects to radio surveys is ongoing; for instance \cite{Hill:2022liq} consider the relative strength of the galaxy intrinsic alignment signals in optical and radio surveys.

One significant advantage of radio weak lensing surveys is that many of these systematic effects are radically different in radio surveys to optical ones. This means that when we do cosmology with cross-correlations of cosmic shear in the two different experiments, then any systematics which are additive to the individual estimates will disappear, and any which are multiplicative can be `self-calibrated' out by using the combination of two auto- and one cross-correlations, as described in detail in \cite{Camera:2016owj,Ingrao:2025ip}. 

\textcolor{black}{Another direction that is promising for increasing the cosmological constraining power and disentangling systematics is to consider beyond 2-point weak lensing statistics \citep{Patton2017,Porqueres2022,Euclid_HOWLS2023,Zuercher2023}. The improvement relies on extracting additional non-Gaussian information, as discussed in the chapter on intensity mapping with statistics beyond the power spectrum \review{\citep{Majumdar01.2026.SKA}}. Some statistics admit a theoretical modelling, such as density-split statistics  \citep{Friedrich_2018,Gruen2018,Burger2023}, related moments or probability distributions of the weak lensing convergence \citep{Boyle2021,Castiblanco2024} and aperture mass \citep{Barthelemy2021,Heydenreich2023,Burger2024}, or integrated 3-point statistics \citep{Halder2021,Gong2023,Gebauer2025}. Joint one-point probability distributions \citep{Friedrich2025}  can serve as a non-Gaussian analogues of auto- and cross-power spectra between an optical and radio survey.
A wider range of scales and statistics such as critical points including voids and peaks \citep{HarnoisDeraps2024} and Minkowski functionals \citep{Armijo2025} are accessible through simulation-based inference \citep{Giblin2023,gatti2024,Marques2024}.}

External data from CMB lensing also provides a third source of lensing information with which to calibrate systematics, with \cite{Kalaja:2024tsk} showing how it can be used to infer the redshift distribution of the radio population; information which otherwise can only be determined for a fraction of sources which have cross-matches at other wavelengths.

\subsection{Forecast for Measurements with AA4}
\label{sec:aa4shear}
Here we show forecasts for tomographic cosmic shear cosmology with SKA-Mid AA4, updating previous forecasts from \cite{Harrison:2016stv,Camera:2016owj,SKA:2018ckk} to consider cross-correlations between SKA-Mid AA4 and a a Stage-IV optical/nIR survey (which we refer to as \ofour) representative of \emph{Euclid} or the Vera C. Rubin Observatory LSST.
\subsubsection{Assumed Surveys}
\label{sec:survey}
The power of a cosmic shear survey in constraining cosmological structure growth can be largely estimated by considering:
\begin{itemize}
    \item the source number density of galaxies with measurable shapes: $n_{\rm gal}\,$arcmin$^{-2}$
    \item the redshift distribution of these sources $n(z)$, which may be sub-divided into a number $N_{\rm bin}$ of tomographic bins. These may become `blurred' in redshift space due to the inherent uncertainties in photometric redshift information, which is typically given as $\sigma_z (1 + z)$.
    \item the fraction of sky covered by the survey $f_{\rm sky}$
    \item the intrinsic random dispersion in galaxy ellipticities which must be averaged over in order to use their shape as an estimate of the cosmic shear. This is usually referred to as `shape noise' $\sigma_\epsilon$
\end{itemize}
In \cref{tab:surveys} we show the values assumed for these quantities in our forecasts. \review{For SKA-Mid AA4 such a survey is achievable with $\sim\!10,000\,$hours of observing time allowing integration to a depth of $2.9\,\mu\mathrm{Jy/beam}$ using the lowest third of Band 2 ($0.95$-$1.22\,\mathrm{GHz}$). These values were obtained by \cite{Bonaldi:2016lbd} using sensitivity curves from \cite{braun2019anticipatedperformancesquarekilometre}, which remain consistent with the current SKAO sensitivity calculator\footnote{\url{https://sensitivity-calculator.skao.int/}}. We note that such a deep, wide area Band 2 survey will likely have additional legacy value across many science areas, following in the footsteps of the likes of FIRST and NVSS.} 

For the redshift distribution of sources $n(z)$ we use the well-known `Smail' functional form:
\begin{equation}
    n(z) = \bar n\,\left( \frac{z}{z_{\rm m}/\sqrt2} \right)^2\,\exp\left[-\left( \frac{z}{z_{\rm m}/\sqrt2} \right)^{3/2}\right]\;,
	\label{eq:redshift distribution}
\end{equation}
wherein $z_{\rm m}$ is the median redshift of the sources. For more details on the modelling of these surveys in the context of radio weak lensing we refer the reader to \cite{SKA:2018ckk,Ingrao:2025ip} where the radio survey is referred to as the `SKA1 Medium-Deep'. This nominal survey is also of interest to other science cases within this volume, including \cite{Asorey01.2026.SKA,Harrison01.2026.SKA,Bertacca01.2026.SKA}.
\begin{table}
	\centering	
	\caption{Survey parameters used in \cref{sec:aa4shear} for forecasting cosmic shear cosmological constraints with SKA-Mid and the optical Stage-IV (\ofour) surveys. Please see text for a description of columns.}
	\vspace{-0.2cm}
	\begin{tabular}{cccccc}
            \hline
            Name & $A_{\rm sky}\, [\mathrm{deg}^{2}]$ & $n_{\rm gal} \, [\mathrm{arcmin}^{-2}]$ & $N_{\rm bins}$ & $z_{\rm m}$ & $\sigma_\epsilon$\\
		\hline
            SKA-Mid AA4 & 5,000 & 2.7 & 5 & 1.1 & 0.3 \\
            O$_{\rm IV}$ & 20,000 & 30 & 10 & 0.9 & 0.3 \\
            \hline
		\hline
	\end{tabular}
	\label{tab:surveys}
\end{table}

\subsubsection{Forecasting Procedure}
\label{sec:systematics:forecasts:results}
\begin{table}
	\centering	
	\caption{	
        The parameters and priors used in the model specification within a $\Lambda$CDM cosmology, including the randomised systematics parameters $X_i$. Fiducial values are used for simulations and initialisation of inference chains. \review{Priors are given as Uniform distributions $\mathcal{U} [\mathrm{min}, \mathrm{max}]$}. Unlisted other cosmological parameters and model choices are fixed to their default values in \camb v1.3.5 \citep{Lewis:1999bs}}
	\vspace{-0.2cm}
	\begin{tabular}{ccc}
            \hline
		Parameter & Fiducial & Prior \\\hline
		\multicolumn{3}{l}{\textbf{Cosmology Sampled}}  \\
		$\Omega_{\rm c}$ & 0.27  & $\mathcal{U}[0.19, 0.35]$ \\ 
        $\sigma_8$ & 0.83  & $\mathcal{U}[0.69, 0.97]$ \\ 
        $\Omega_{\rm b}$ & 0.0.045  & $\mathcal{U}[0.01, 0.08]$ \\ 
		$h$  & 0.67 & $\mathcal{U}[0.40, 0.94]$   \\
        $n_s$  & 0.966 & $\mathcal{U}[0.90, 1.02]$   \\
        $w_0$  & -1.0 & $\mathcal{U}[-1.5, -0.5]$   \\
        $w_a$  & 0.0 & $\mathcal{U}[-2.0, 2.0]$   \\
		\hline

            \multicolumn{3}{l}{\textbf{Systematics Sampled} } 	 \\
            $X_i$ & $<0.1$ & $\mathcal{U}[-0.05, 0.05]$ \\
            \hline
  
		
		
		\hline
	\end{tabular}
	\label{tab:priors}
\end{table}
To forecast the performance of SKAO cosmic shear cosmology we simulate cosmic shear power spectrum at the fiducial \lcdm cosmology given in \cref{tab:priors} \review{with stated Uniform priors}. We assume an \ska survey as described in \cref{sec:survey}. For qualitative comparison we also consider a Stage-III optical/nIR survey (representative of DES, KiDS or HSC) and a Stage-IV optical/nIR survey representative of \emph{Euclid} or the Vera C. Rubin Observatory LSST.


The theoretical power spectra are modelled using \camb \citep{Lewis:1999bs} as the Boltzmann solver, with \halofit \citep[the version of][]{Mead:2020vgs} for the non-linear matter power spectrum, and Limber integrals carried out using \ccl \citep{LSSTDarkEnergyScience:2018yem}. We then assume a Gaussian likelihood in the power spectrum $C_\ell$s and an analyitc Gaussian `Knox' covariance. The \texttt{emcee} package \citep{2013PASP..125..306F} is then used to estimate the posterior constraints on cosmological and nuisance model parameters.

We use angular scales in the power spectra $2 \leq \ell \leq 2000$, intended as a reasonable choice in line with \cite{Euclid:2019clj}, representing a scenario in which non-linear effects on the matter power spectrum can be accurately modelled.

For cosmic shear nuisance parameters we follow the approach of \cite{Camera:2016owj} in applying random multiplicative systematics to the tomographic shear power auto- and cross-power spectra between the different tomographic bins in the optical and radio experiments. These take the form of a template matching the cosmic shear power spectrum, with an amplitude for each tomographic bin $X_i$. The values of these $X_i$ are randomly sampled \review{from a Uniform distribution} between $[-5\%, 5\%]$ and are then marginalised over alongside cosmological parameters in the posterior estimation. \review{Typical values of multiplicative shear bias parameters for current optical weak lensing surveys are $\sim\!0.5\%$ \citep[e.g.][]{DES:2026mkc}, meaning these represent conservative estimates for the amount of calibration which may be required for SKA radio surveys.}
\subsubsection{Results}
\begin{figure}[h]
    \centering
	\includegraphics[width=0.5\columnwidth]{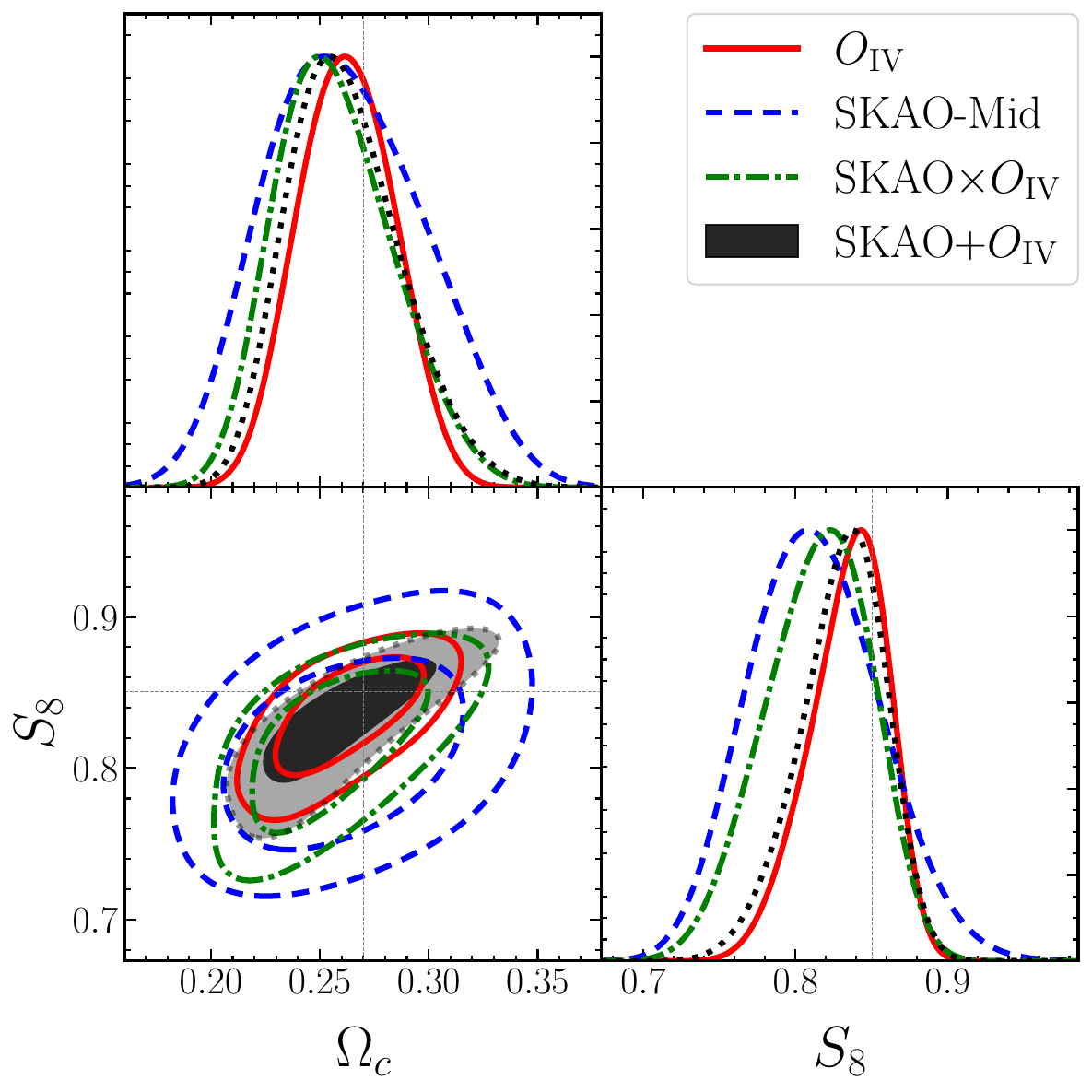}
    \caption{The constraining power on $S_8$ parameters from the cosmic shear surveys described in the text. Dashed blue is that for the SKA-Mid AA4 survey auto-correlations, solid red for the \ofour optical/nIR Stage-IV survey auto-correlations, dash-dotted green represents their cross-correlation (which will remove additive systematics by construction), and filled contours show the combination of the two auto- plus one cross-correlation from the two surveys, which can be used to remove both additive and multiplicative systematics following the approach of \cite{Camera:2016owj,Ingrao:2025ip}}.
    \label{fig:results}
\end{figure}

The results of the forecasting exercise are shown in \cref{fig:results}. As can be seen \emph{under our general and flexible model for additive and multiplicative systematics} SKA-Mid AA4 alone is capable of measuring the $S_8$ parameter to $5 \%$ precision, compared to $2 \%$ for the \ofour survey. Note these are not directly comparable to current real data constraints on $S_8$ from e.g. DES-Y3 because of both this more general treatment of systematics and the inclusion of marginalisation over $w_0 - w_a$ parameters in our forecasts. The cross-correlation, \emph{which will by construction remove additive systematics which may be present in either experiment}, also provides a $4 \%$ constraint. Finally, the full combination of all auto- and cross-spectra \emph{which can mitigate both additive and multiplicative systematics} through self-calibration provides a systematic-free measurement of $S_8$ at $3 \%$ precision. Such precision with guaranteed accuracy could help in definitively identifying the physical origin of dark matter and dark energy which drive how the value of $S_8$ evolves with redshift and physical scale, two of the most important endeavours in modern cosmology.

\begin{table}[ht]
    \centering
    \caption{Asymmetric 68\% credible intervals shown as $x^{+\Delta_+}_{-\Delta_-}$ for $\Omega_m$–$S_8$ constraints with the cosmic shear surveys considered here.}
    \begin{tabular}{lccc}
    \hline
    Dataset & $\sigma_8$ & $S_8$ & $\Omega_m$ \\
    \hline
    SKAO-Mid & $0.8737^{+0.0620}_{-0.0666}$ & $0.8106^{+0.0432}_{-0.0460}$ & $0.2575^{+0.0355}_{-0.0444}$ \\
    \ofour & $0.8956^{+0.0299}_{-0.0288}$ & $0.8395^{+0.0293}_{-0.0226}$ & $0.2621^{+0.0241}_{-0.0226}$ \\
    SKAO$\times$\ofour & $0.8841^{+0.0374}_{-0.0408}$ & $0.8187^{+0.0398}_{-0.0334}$ & $0.2531^{+0.0244}_{-0.0314}$ \\
    SKAO$+$\ofour & $0.8948^{+0.0278}_{-0.0270}$ & $0.8344^{+0.0303}_{-0.0258}$ & $0.2578^{+0.0231}_{-0.0298}$ \\
    \hline
    \hline
    \end{tabular}
\label{tab:omega_m_S8_results}
\end{table}

\subsection{Impact of Potential SKA Enhancements}
\label{sec:ska2}
Previous works have all considered cosmic shear surveys with SKA2 (or `the full SKA') with the expectation that this configuration has both higher resolution (by factors $\sim 2$) and greater sensitivity (by factors $\sim 10$) enabling measurement of shapes of orders of magnitude more sources with consequential increase in cosmological constraining power. Here, for completeness we report the outcome of forecasts using the machinery from \cref{sec:aa4shear} but with an SKA2 configuration. We find the constraining power on $\sigma_8$ improves from $8\%$ to $1.5\%$, in line with expectations from \citealt{Harrison:2016stv,SKA:2018ckk}, and to the point at which SKAO exceeds the power of \ofour surveys and becomes the leading cosmic shear survey.

We emphasise here that other expansions of the SKA-Mid telescope intermediate which are between AA4 and SKA2 will have correspondingly intermediate effects in increased cosmological constraining power, if they allow the shape measurement (through detection and resolving) of larger numbers of high redshift star-forming galaxies. In particular, the preference for cosmic shear surveys would be for baseline distributions which create a PSF which best approximates a matched-filter for such galaxies; likely a $\sim1\,$arcsec exponential form, as explored in detail in Section 6 of \cite{Harrison:2020zsv}. This science case would therefore benefit significantly from expansions of SKA-Mid which increase the baseline density in the $70$-$100\,$km range, as also demonstrated in \cref{fig:aa4_aastar} of this chapter.

\section{Weak Lensing with Other Radio Information}
\label{sec:systematics}
In addition to being able to mitigate additive and multiplicative systematics through cross-correlations an self-calibration as discussed above, radio surveys also continue uniquely available information which can significantly improve the precision and accuracy of cosmic shear (and subsequent cosmological constraints). `Shape noise' in weak lensing analyses arises from the degeneracy between cosmological weak gravitational lensing and a galaxies intrinsic ellipticity. This necessitates the large source number density of resolved galaxies in cosmic shear. The shapes of these are effectively averaged, with the intrinsic ellipticity of multiple sources expected to converge to zero and the residual being the shear estimate. However, averaging over finite numbers of galaxies with an intrinsic ellipticity dispersion acts as a statistical floor on the size of shear signal which can be distinguished. Both galaxy polarisation and kinematics may be used to estimate the un-lensed position angle of sources, which has has been shown to be comparable in cosmological constraining power to the full ellipticity information \citep{Whittaker:2013cga,Whittaker:2015fma}.

In addition to this statistical limitation, galaxy shapes in reality have intrinsic alignments (IA) as a result of their formation processes, resulting in a systematic contamination to the inferred gravitational lensing signal. Modelling, nulling, or otherwise accounting for this contamination is a focus of considerable work within the weak lensing community \citep[for an extensive discussion see][]{Chisari:2025gsy}. By providing information on the un-lensed intrinsic shape of sources, radio surveys can mitigate this systematic and recover the power of surveys which may otherwise be lost to marginalisation over IA nuisance parameters.
\subsection{Polarisation}
Because the integrated polarisation angle of a source is not changed by gravitational lensing, and is expected to correlate with the intrinsic position angle of a galaxy, we can use it as a measure of the pre-lensing intrinsic shape of weak lensing sources. This addresses both the shape noise and intrinsic alignment issues. \cite{Brown:2010rr} first proposed this technique, demonstrating its ability to mitigate intrinsic alignment contamination at a level appropraite for Stage-IV surveys, predicated on reasonable polarisation fractions and polarisation alignment. \cite{Thomas:2016xhb} show how polarisation can also be used to extract information on lensing-induced rotation of galaxy images, which has the potential to detect both exotic \citep[but highly topical e.g.][]{Diego-Palazuelos:2025dmh} birefringence physics, and otherwise control systematic errors. A potential equivalent to `shape noise' comes from stochastic mis-alignments between galaxies' polarisation and position angles; \cite{Zhou:2025bvz} investigate this using the IllustrisTNG simulation, finding some degradation of the alignment with increasing redshift, but showing polarisation information can still contribute competitively in estimating intrinsic shapes for weak lensing.
\subsection{Kinematics}
Similar to polarisation, the achromatic nature of gravitational lensing means that the velocity field of a source can be used as a tracer of its intrinsic shape. Un-resolved velocity measurements can be combined with the Tully-Fisher (TF) relation to provide the position angle misalignment to be used as a cosmic shear estimate. Following the initial suggestion by \cite{Huff:2013dha} this TF weak lensing has begun to be observed in optical data \citep{S:2024vig} and is expected to be able to enhance the power of weak lensing in Roman Space Telescope surveys \citep{Xu:2022lne}. The use of 21cm neutral hydrogen line measurements with SKAO has also been proposed by \cite{Huang:2025kme} who show it increases the cosmological constraining power of a sample which has both TF and traditional imaging weak lensing measurements together.

Furthermore, where resolved kinematics are available, \cite{Morales:2006fq} has shown how this can be used to estimate the shear using individual objects, which has subsequently been measured by \cite{2020MNRAS.499.4591G} and argued to produce gains by a factor 2-6 in constraining power of weak lensing surveys \citep{DiGiorgio:2021egh}, although detailed forecasts for SKAO neutral hydrogen measurements do not yet exist.

\subsection{21cm Intensity Maps}
In principle it is also possible to follow well-established techniques from CMB observations for inferring the lensing of a nearly-Gaussian intensity field \cite[as reviewed in][]{Lewis:2006fu}. Neutral Hydrogen 21cm intensity maps, as discussed at length elsewhere in this volume, can then trace large scale lensing signals with high redshift resolution. \cite{Pourtsidou:2015mia} showed the detectability of this signal at SNR$\sim10$ with an SKA-Mid AA4-like configuration. 
\cite{LozanoTorres:2022mci} show how the signal may then be used for full three dimensional weak lensing analysis, which can contain significantly more information than the tomographic approaches considered above (and in almost all other weak lensing surveys).

\section{Conclusions}
In this chapter we have discussed the current observational status and potential future payoff from conducting cosmic shear cosmology surveys in the radio with SKA-Mid AA4. The utility of SKA surveys for cosmic shear has long been established \citep{Kamionkowski:1997mp,Blake:2004pb,Brown:2015ucq}, in particular their ability to provide critical robustness to experimental and astrophysical systematics when complementing front-line optical/nIR cosmic shear surveys.

In the ten years since the preceding SKA Science Book, steady observational progress has been made, both in extracting the available galaxy shape information from legacy surveys (in FIRST \citealt{Demetroullas:2015gsa,2018MNRAS.473..937D}, COSMOS \citealt{Tunbridge:2016oio,Hillier:2018gdr} and LOFAR \citealt{Liu:2025sbv}) and in bespoke surveys (SuperCLASS \citealt{2020MNRAS.495.1706B,Harrison:2020zsv}).

This observational progress has also lead to a much greater understanding of the necessary methodological developments. It was already understood that `standard' deconvolution imaging techniques would be inadequate for cosmic shear \citep{Patel:2015cra,Harrison:2015lba} and much effort has gone into developing alternatives tailored for measuring galaxy ellipticity to high precision and accuracy both in the image plane \citep{Harrison:2020zsv,2025A&A...696A.216T} and the $uv$-plane \citep{Rivi:2016ccv,Rivi:2017rym,Rivi:2018gnm,Rivi:2022lha}.

In this Chapter we have shown updates to previous forecasts, which considered previous combinations of `SKA1' with Stage-III optical surveys currently approaching completion (DES, KiDS, HSC) and of `SKA2'/`the full SKA' with Stage-IV optical surveys which are now approaching science verification stages (Vera C. Rubin LSST, \emph{Euclid}). Though the constraining power of SKA-Mid AA4 will be at a Stage-III-like level, we have seen how combinations with a Stage-IV optical survey can nonetheless provide interesting constraints on cosmological parameters whilst gaining critical systematics robustness. Such systematics robustness remains a compelling feature, with the presence or absence of an `$S_8$ tension' in existing weak lensing data still unclear.

In \cref{sec:systematics} we have also shown how the period since the previous Science Book has led to exciting new developments in the use of additional information from polarisation and kinematics for cosmic shear cosmology surveys. This highly promising new observable has begun to be detected \citep{S:2024vig}, and begun to be explored in radio surveys \citep{Huang:2025kme}. Further work is required to fully explore the power of SKA-Mid AA4, and to make concerted efforts detecting the signal in pathfinders and precursors.

Finally, in \cref{fig:aa4_aastar} we have shown the critical importance of the long baselines included in the AA4 configuration for this science case. Without these baselines, the AA$^*$ configuration is not capable of measuring the galaxy shape information necessary for detecting a useful cosmological signal. Should AA4 be acheived and built upon, we have also shown in \cref{sec:ska2} how the inclusion of more longer baselines in an `SKA2' configuration leads to a transformational impact on the ability to do cosmic shear surveys which are truly world-leading.

\section*{Author Ordering}
Authors are listed in two tiers. The first tier is in order of contribution. The second tier is listed in alphabetical order.

\section*{Acknowledgements}
We thank members of the SKAO Cosmology SWG and the Weak Lensing Focus Group for helpful discussions over many years. RI and SC acknowledge support from the Italian Ministry of University and Research (\textsc{mur}), PRIN 2022 `EXSKALIBUR – Euclid-Cross-SKA: Likelihood Inference Building for Universe's Research', Grant No.\ 20222BBYB9, CUP D53D2300252 0006, from the Italian Ministry of Foreign Affairs and International
Cooperation (\textsc{maeci}), Grant No.\ ZA23GR03, and from the European Union -- Next Generation EU. We acknowledge the support of the Supercomputing Wales project, which is part-funded by the European Regional Development Fund (ERDF) via Welsh Government.

\bibliographystyle{abbrvnat-maxbibnames4}
\bibliography{chapter_newstyle} 

\end{document}